\documentclass[12pt,epsf]{article}
\usepackage{enumerate,amsmath,amscd,amssymb,graphicx 
}

\makeatletter
\@addtoreset{equation}{section}

\renewcommand{\thefootnote}{\fnsymbol{footnote}}
\makeatother

\begin{document}

\newcommand {\beq}{\begin{eqnarray}}
\newcommand {\eeq}{\end{eqnarray}}
\newcommand {\non}{\nonumber\\}
\newcommand {\eq}[1]{\label {eq.#1}}
\newcommand {\defeq}{\stackrel{\rm def}{=}}
\newcommand {\gto}{\stackrel{g}{\to}}
\newcommand {\hto}{\stackrel{h}{\to}}
\newcommand {\1}[1]{\frac{1}{#1}}
\newcommand {\2}[1]{\frac{i}{#1}}
\newcommand {\thb}{\bar{\theta}}
\newcommand {\ps}{\psi}
\newcommand {\psb}{\bar{\psi}}

\newcommand {\ph}{\varphi}
\newcommand {\phs}[1]{\varphi^{*#1}}
\newcommand {\sig}{\sigma}
\newcommand {\sigb}{\bar{\sigma}}
\newcommand {\Ph}{\Phi}
\newcommand {\Phd}{\Phi^{\dagger}}
\newcommand {\Sig}{\Sigma}
\newcommand {\Phm}{{\mit\Phi}}
\newcommand {\eps}{\varepsilon}
\newcommand {\del}{\partial}
\newcommand {\dagg}{^{\dagger}}
\newcommand {\pri}{^{\prime}}
\newcommand {\prip}{^{\prime\prime}}
\newcommand {\pripp}{^{\prime\prime\prime}}
\newcommand {\prippp}{^{\prime\prime\prime\prime}}

\newcommand {\pripppp}{^{\prime\prime\prime\prime\prime}}
\newcommand {\delb}{\bar{\partial}}
\newcommand {\zb}{\bar{z}}
\newcommand {\mub}{\bar{\mu}}
\newcommand {\nub}{\bar{\nu}}
\newcommand {\lam}{\lambda}
\newcommand {\lamb}{\bar{\lambda}}
\newcommand {\kap}{\kappa}
\newcommand {\kapb}{\bar{\kappa}}
\newcommand {\xib}{\bar{\xi}}
\newcommand {\ep}{\epsilon}
\newcommand {\epb}{\bar{\epsilon}}
\newcommand {\Ga}{\Gamma}
\newcommand {\rhob}{\bar{\rho}}
\newcommand {\etab}{\bar{\eta}}
\newcommand {\chib}{\bar{\chi}}
\newcommand {\tht}{\tilde{\th}}
\newcommand {\zbasis}[1]{\del/\del z^{#1}}
\newcommand {\zbbasis}[1]{\del/\del \bar{z}^{#1}}
\newcommand {\vecv}{\vec{v}^{\, \prime}}
\newcommand {\vecvd}{\vec{v}^{\, \prime \dagger}}
\newcommand {\vecvs}{\vec{v}^{\, \prime *}}
\newcommand {\alpht}{\tilde{\alpha}}
\newcommand {\xipd}{\xi^{\prime\dagger}}
\newcommand {\pris}{^{\prime *}}
\newcommand {\prid}{^{\prime \dagger}}
\newcommand {\Jto}{\stackrel{J}{\to}}
\newcommand {\vprid}{v^{\prime 2}}
\newcommand {\vpriq}{v^{\prime 4}}
\newcommand {\vt}{\tilde{v}}
\newcommand {\vecvt}{\vec{\tilde{v}}}
\newcommand {\vecpht}{\vec{\tilde{\phi}}}
\newcommand {\pht}{\tilde{\phi}}
\newcommand {\goto}{\stackrel{g_0}{\to}}
\newcommand {\tr}{{\rm tr}\,}
\newcommand {\GC}{G^{\bf C}}
\newcommand {\HC}{H^{\bf C}}
\newcommand{\vs}[1]{\vspace{#1 mm}}
\newcommand{\hs}[1]{\hspace{#1 mm}}
\newcommand{\al}{\alpha}
\newcommand{\be}{\beta}
\newcommand{\Lam}{\Lambda}

\newcommand{\kahler}{K\"ahler }
\newcommand{\con}[1]{{\Gamma^{#1}}}
\newcommand {\dellr}{\stackrel{\leftrightarrow}{\partial}}

\thispagestyle{empty}
\begin{flushright}
TIT/HEP--520 \\
hep-th/0405161 \\
May, 2004 \\
\end{flushright}
\vspace{3mm}
\begin{center} 
{\large {\bf
Effective Theory on 
Non-Abelian Vortices in Six Dimensions 
}}
\vspace{5mm}

  {\bf 
  Minoru~Eto}
\footnote{\it  e-mail address: 
meto@th.phys.titech.ac.jp
},~
  {\bf 
  Muneto~Nitta}
\footnote{\it  e-mail address: 
nitta@th.phys.titech.ac.jp
}  
~and~~ {\bf 
Norisuke~Sakai}
\footnote{\it  e-mail address: 
nsakai@th.phys.titech.ac.jp
}

\vskip 1.5em

{ \it 
  Department of Physics, Tokyo Institute of 
Technology \\
Tokyo 152-8551, JAPAN 
   }
\vspace{20mm}

{\bf Abstract}\\[5mm]
{\parbox{13cm}{\hspace{5mm}
Non-Abelian vortices in six spacetime dimensions 
are obtained for a supersymmetric $U(N)$ 
gauge theory with $N$ hypermultiplets in the 
fundamental representation. 
Massless (moduli) fields are identified and classified into 
Nambu-Goldstone and quasi-Nambu-Goldstone fields. 
Effective gauge theories for the moduli fields are 
constructed 
on the four-dimensional world volume of vortices. 
A systematic method to obtain 
the most general form of the effective Lagrangian 
consistent with symmetry is proposed. 
The moduli space for the multi-vortices 
is found to be a vector bundle over 
the complex Grassmann manifold. 

}}
\end{center}
\vfill
\newpage
\setcounter{page}{1}
\setcounter{footnote}{0}
\renewcommand{\thefootnote}{\arabic{footnote}}

\section{Introduction}

The brane-world scenario~\cite{HoravaWitten,LED,RandallSundrum} 
has attracted much attention in recent years. 
These models with extra dimensions require 
some topological defects in higher dimensional spacetime. 
It is often advantageous to consider supersymmetric 
(SUSY) models 
in order to construct such topological defects. 
The topological defects typically break part of the 
original SUSY~\cite{WittenOlive} 
leaving the ${\cal N}=1$ SUSY in 
four-dimensional world volume of the topological defect. 
In this way we may be able to obtain realistic unified 
models of the type of the minimal SUSY standard 
model~\cite{DGSW}. 

Since standard model particles should be realized 
as low-energy fluctuations localized on topological 
defects such as walls and vortices, it is important to study 
modes which are massless or nearly massless 
compared to the mass scale of the 
topological defects. 
When we consider a solution of the equations of motion as 
a possible background corresponding to the topological defect, 
it can often contain parameters. 
These parameters are called moduli and represent 
possible deformations of the 
background solution without costing energy. 
Therefore one can promote the parameters into fields 
on the world volume of the topological defect~\cite{Ma}. 
These fields are massless and are called moduli fields. 
The moduli fields describe the low-energy dynamics 
of the topological defect. 
Some of the massless fields may originate from 
spontaneously broken global symmetries and 
are called Nambu-Goldstone particles. 
Number of the Nambu-Goldstone particles are given by 
the number of broken symmetry generators including 
the spacetime symmetry generators. 

Low-energy interactions of the Nambu-Goldstone particles 
are severely constrained by the low-energy theorems 
which may be obtained by the method of nonlinear 
realization~\cite{Coleman:sm}. 
On the other hand, there may be massless particles which 
do not correspond to the spontaneously broken generators. 
In the case of SUSY theories where ${\cal N}=1$ 
SUSY is maintained in four-dimensions, 
scalar particles have to form complex fields 
in order to form a chiral scalar 
field. 
If a Nambu-Goldstone particle does not 
form a complex scalar with another Nambu-Goldstone 
scalar, it still has to accompany a massless scalar 
to form a chiral scalar field. 
Such massless scalars do not correspond to broken 
symmetry generators and are required to exist only 
because of SUSY. 
These massless scalars are called 
quasi-Nambu-Goldstone bosons~\cite{BLPY}--\cite{HNOO}, 
and will acquire masses 
once SUSY is broken. 
For a model building, it is extremely useful to obtain 
Nambu-Goldstone particles as well as 
quasi-Nambu-Goldstone particles to obtain 
particles in the low-energy effective theories. 

Walls are the 
simplest topological defects with single extra 
dimension. 
Walls with eight SUSY, whose world-volume 
theories preserve four SUSY, were discussed in \cite{DW}.
If we have two extra dimensions, we need to have topological 
defects such as vortices. 
Vortex is a typical topological defect with codimension two. 
In the pioneering work of 
Abrikosov~\cite{Abrikosov:1956sx}, and Nielsen and 
Olsen~\cite{Nielsen:cs} they have worked out a 
vortex in the $U(1)$ gauge theory, 
which is called the ANO 
vortex.  
Vortices with higher vorticities~\cite{dVegaSchaposnik} 
and vortices in higher dimensions~\cite{HLP} 
have also been constructed. 
In the brane-world scenario, vortices are especially useful 
to localize gauge fields using the warped 
compactifications~\cite{Gravity}. 
More recently, 
monopoles in the Higgs phase 
have been 
found~\cite{Hanany:2003hp}--\cite{HT2} 
in the three- or four-dimensional 
SUSY gauge theories 
with eight SUSY. 
They can be static solutions when they are attached to 
non-Abelian vortices at both sides~\cite{Tong:2003pz}. 
These non-Abelian vortex solutions 
are obtained by deforming the 
$U(1)$ ANO vortex embedded 
into the SUSY non-Abelian gauge theories.   
It has been observed that the non-Abelian vortices have 
interesting characteristics compared to Abelian 
vortices. 
Non-Abelian monopoles in the Higgs phase are also discussed 
in the context of monopole confinement by the $Z_k$ vortex
in the deformed ${\cal N}=2$ and ${\cal N}=4$ SUSY Yang-Mills 
theories\cite{Kneipp}.

The number of moduli fields can be counted by means of 
index theorems for many topological defects, such as 
vortices~\cite{Weinberg}. 
More recently,  new methods have been proposed to 
construct effective theories of moduli fields 
on the world volume of the topological defect, 
such as vortices or instantons~\cite{Hanany:2003hp,HT2}.  
One of the methods uses a field theoretical 
Ansatz~\cite{Auzzi:2003fs}, 
which becomes simple for a particular case where the $U(1)$ 
gauge coupling $e$ is related to the 
$SU(N)$ gauge coupling $g$ by $g^2=Ne^2$~\cite{Hanany:2003hp}. 
The other method is inspired by brane constructions 
and gives a gauge theory on the world volume. 
Both of them have been applied to the three-dimensional 
$U(N)$ gauge theories 
with eight SUSY, leading to gauge theories 
with four SUSY in the two-dimensional world volume 
of the vortices~\cite{Hanany:2003hp}. 
Although they have considered a particular case 
between coupling constants, 
the results are expected to apply 
to other cases (with some deformations of moduli metric 
and so on). 
Since there are ambiguities to define the moduli fields, 
one can have different metric of moduli fields in various 
approaches for low-energy dynamics of topological 
defects such as vortices. 
In particular it has been noted that the moduli metric 
obtained by using the method of Manton~\cite{Ma} 
is different from that deduced from the world-volume gauge 
theory inspired by brane constructions~\cite{Hanany:2003hp}. 

The purpose of this paper is to discuss non-Abelian BPS 
vortex solutions in a six-dimensional $U(N)$ 
SUSY gauge theory with $N$ hypermultiplets in the fundamental 
representation and to work out the effective 
field theories for their moduli fields on the 
four-dimensional world volume, so that it may be 
useful for a model building in the brane-world 
scenario. 
We work out BPS equations for multi-vortex 
solutions and construct their moduli space. 
We succeed to distinguish the Nambu-Goldstone fields 
and quasi-Nambu-Goldstone fields. 
We also obtain general form of the effective field theories 
for the moduli fields in the spirit of the world-volume gauge 
theories 
inspired by brane constructions~\cite{Hanany:2003hp}. 
Since the effective theories on the BPS vortices have 
four SUSY, we obtain K\"ahler quotient construction 
for the Grassmann manifold as the moduli space. 
We also work out possible forms of the metric 
consistent with the symmetry requirement of the 
low-energy effective theory. 
This will hopefully explain the 
discrepancies~\cite{Hanany:2003hp,Ma} noted 
previously between two different methods to derive 
the low-energy effective theory. 
Similarly the deformations of the moduli metric 
in the case of $g^2\not = Ne^2$ may be also incorporated 
by this method. 

In sect.\ref{sc:Lagrangian}, 
the six-dimensional Lagrangian with eight SUSY 
is constructed using the superfield representing four 
SUSY, at the cost of sacrificing the manifest 
Lorentz invariance in six dimensions. 
In sect.\ref{sc:NonAbelianVortices}, 
non-Abelian vortices are constructed in six dimensions. 
In sect.\ref{sc:EffectiveTheory}, 
effective theory on the world volume of vortices is 
constructed. 
In sect.\ref{sc:deformation}, 
general form of deformations of vortex moduli space 
is studied. 
Sect.\ref{sc:discussion} 
is devoted to a discussion of possible uses of 
our results in the brane-world context.

\section{$D=6$ Lagrangian in 
4 SUSY Superspace}
\label{sc:Lagrangian}
We start with six-dimensional 
${\cal N}=1$ (i.e. eight supercharges) supersymmetric 
(SUSY) gauge theory 
with the $U(1) \times SU(N_{\rm C})$ gauge group 
and $N_{\rm F}$ hypermultiplets with the same (unit) 
$U(1)$ charges. 
We are especially interested in the case of 
$N_{\rm C} = N_{\rm F} \equiv N$. 
We will study non-Abelian vortices 
in six-dimensions, which lie in the $x^5$-$x^6$ plane 
and have four-dimensional world-volume 
extending from $x^0$ to $x^3$ 
applicable to the brane-world scenario.
For this purpose, 
it is useful to express the $D=6$ Lagrangian 
in terms of 4 SUSY superfield 
formalism discussed in \cite{Arkani-Hamed:2001tb}. 
At the cost of sacrificing 
the manifest $D=6$ Lorentz invariance, 
we can use this superfield formalism, 
maintaining the $D=4$ Lorentz invariance. 
We first review their formalism briefly in this section.

Each $D=6$ supermultiplet is decomposed into 
a set of 4 SUSY superfields (with dependence of 
$x^5$ and $x^6$ coordinates) as follows.
The $U(1)$ vector multiplet is decomposed 
into $(V,\Phi)$ with $V$ and $\Phi$ vector and chiral 
superfields, respectively, in 4 SUSY superfield formalism.  
A set of $N$ by $N$ matrix superfields 
$(\hat V,\hat\Phi)$ constitutes 
$D=6$ $SU(N)$ vector multiplets.
Here a hat denotes 
a matrix in the fundamental representation of 
$SU(N)$ which can be decomposed in terms of 
basis $T^A$ 
\begin{eqnarray}
\hat X^a{_b} = 
\sum_{A=1}^{N^2 -1} X^A \left(T^A\right)^a{_b},
 \nonumber
\end{eqnarray}
where $A$ 
is the adjoint index running 
$1, \cdots,N^2-1$, 
and $a,b$ are 
indices of the fundamental representation 
running $1,\cdots,N_{\rm C}=N$. 
The $N_{\rm F}$ hypermultiplets in the fundamental representation 
are decomposed into $(Q_i,\tilde Q^i)$ 
($i=1,\cdots,N_{\rm F}=N$) 
with $Q_i$, $\tilde Q^i$ chiral superfields belonging 
to the fundamental and the anti-fundamental 
representations of $SU(N)$, respectively, 
with the opposite $U(1)$ charges. 
Here we have denoted $Q_i$ and $\tilde Q^i$ by 
column and row vectors whose components are denoted by 
$a$, 
but sometimes we use matrix notations, 
$Q^a{}_i$ and $\tilde Q^i{}_a$. 

Let us denote the six-dimensional indices by 
capital letters $M= 0, 1, 2, 3, 5, 6$, 
and four-dimensional indices 
by small letters $m= 0, 1, 2, 3$. 
We define the covariant derivative 
\begin{equation}
 {\cal D}_M \equiv \partial_M 
 - \frac{i}{2}A_M
 - \frac{i}{2}\hat A_M .
\end{equation}
We can decompose vector fields $A_M, \hat A_M$ 
in six dimensions 
to vector fields $A_m, \hat A_m$ and complex scalar 
fields $\phi, \hat \phi$ 
\begin{eqnarray}
\phi = \frac{A_6 + iA_5}{\sqrt 2},\quad
\hat \phi = \frac{\hat A_6 + i\hat A_5}{\sqrt 2}.
\end{eqnarray}
The field strengths and covariant derivatives in 
extra dimensions can then be rewritten
\footnote{
We denote complex conjugate by 
bar
and hermitian 
conjugate for matrix by ${}^\dagger$. 
} 
\begin{eqnarray}
F_{56} &=& 
\frac{1}{\sqrt 2}\left(\partial \bar\phi 
+ \bar\partial\phi\right),\\
\hat F_{56} &=& 
\frac{1}{\sqrt 2}\left(\partial \hat\phi^\dagger 
+ \bar\partial\hat\phi
+ \frac{1}{\sqrt 2}\left[{\hat\phi}^\dagger,
\hat\phi\right]\right),\\
{\cal D}q^a{_i} &=& 
\partial q^a{_i} - \frac{1}{\sqrt 2}\phi q^a{_i} 
- \frac{1}{\sqrt 2}\hat\phi^a{_b}q^b{_i},\\
\bar {\cal D} \tilde q^{\dagger a}{_i} 
&=& \bar\partial \tilde q^{\dagger a}{_i} 
+ \frac{1}{\sqrt 2}\bar\phi \tilde q^{\dagger a}{_i}
+ \frac{1}{\sqrt 2}{\hat\phi}^{\dagger a}{_b} 
\tilde q^{\dagger b}{_i}.
\end{eqnarray}

Invariance under gauge transformations 
requires a Wess-Zumino-Witten term in general 
gauges~\cite{siegel,Arkani-Hamed:2001tb}.  
In the Wess-Zumino gauge, superfields 
can be expanded in terms of component fields 
as 
\begin{eqnarray}
V(x^m,\theta,\bar\theta;x^5,x^6) &=&
- \theta\sigma^m\bar\theta A_m 
+ i\bar\theta\theta^2\lambda_1 
- i\theta^2\bar\theta\bar\lambda_1
+ \frac{1}{2}\theta^2\bar\theta^2 \left(D - F_{56}\right),
\label{eq:vector-superfield-u1}
\\
\Phi(y^m,\theta;x^5,x^6) &=& \phi 
+ \sqrt 2 \theta\lambda_2 + \theta^2F_\Phi,\\
\hat V(x^m,\theta,\bar\theta;x^5,x^6) &=&
- \theta\sigma^m\bar\theta \hat A_m 
+ i\bar\theta\theta^2\hat\lambda_1 
- i\theta^2\bar\theta\bar{\hat \lambda}_1 
+ \frac{1}{2}\theta^2\bar\theta^2 
\left(\hat D - \hat F_{56}\right),
\label{eq:vector-superfield-un}
\\
\hat\Phi(y^m,\theta;x^5,x^6) 
&=& \hat\phi + \sqrt 2 \theta\hat\lambda_2 
+ \theta^2\hat F_\Phi,\\
Q_i(y^m,\theta;x^5,x^6) &=& q_i + \sqrt 2 \theta \psi_i 
+ \theta^2\left( F_i 
+ \bar{\cal D}\tilde q^\dagger{}_i\right),
\label{eq:hyper-superfield-q1}
\\
\tilde Q^i(y^m,\theta;x^5,x^6) 
&=& \tilde q^i + \sqrt 2 \theta \tilde \psi^i
+ \theta^2\left(\tilde F^i - \bar{\cal D}q^{\dagger i}\right).
\label{eq:hyper-superfield-q2}
\end{eqnarray}
Here $x^m$ $(m=0,1,2,3)$ are the $D=4$ spacetime coordinates 
and $y^m$ are the chiral coordinates defined by 
$y^m \equiv x^m + i\theta \sig^m \thb$ with 
$\theta$ ($\thb$) $D=4$ two-component Weyl (anti-Weyl) 
spinor coordinates in 
the 4 SUSY superspace. 
All component fields in the right hand sides 
depend on $x^m$, $x^5$ and $x^6$. 
Fermions 
$\lam_1$, $\lam_2$, $\hat \lam_1$, $\hat \lam_2$, 
$\psi_i$ and $\tilde \psi^i$ transform as 
$D=4$ two-component Weyl spinors but depend 
on extra coordinates also. 
The gauginos $\lam_1$, $\lam_2$, $\hat \lam_1$, 
and $\hat \lam_2$, together form a symplectic 
Majorana spinor in $6$ dimensions, whereas 
$\psi_i$ and $\tilde \psi^i$ together 
transform as a $D=4$ Dirac spinor. 
Note that 
the  last terms in 
Eqs.(\ref{eq:vector-superfield-u1}), 
(\ref{eq:vector-superfield-un}), 
(\ref{eq:hyper-superfield-q1}), 
and (\ref{eq:hyper-superfield-q2})
are $D$-term and $F$-term components in $V$ and $Q$, 
but are not 
$D=6$ Lorentz scalars 
themselves.
Only after subtracting $F_{56}$, 
$\bar{\cal D}\tilde q^\dagger{}_i$, or 
$\bar{\cal D}q^{\dagger i}$, 
the auxiliary fields 
$D,\hat D,F_i,\tilde F^i$ become genuine 
$D=6$ Lorentz scalars. 
In the last two equations we used 
the holomorphic covariant derivative with respect to 
extra coordinates defined by 
\beq
 {\cal D} \equiv {\cal D}_5 - i{\cal D}_6. 
\eeq

Taking the Wess-Zumino gauge, 
we obtain the Lagrangian~\cite{Arkani-Hamed:2001tb} 
\begin{eqnarray}
{\cal L} &=& {\rm Tr}\bigg[
\frac{1}{4g^2}\left(\int d^2\theta\ 
\hat W^\alpha \hat W_\alpha + \int d^2\bar\theta\ 
{\hat W}_{\dot\alpha}^\dagger 
{\hat W}^{\dot\alpha\dagger}\right)\nonumber\\
&&\qquad+ \frac{1}{g^2}\int d^2\theta d^2\bar\theta\left\{
\left({\hat \Phi}^\dagger 
+ \sqrt 2 \bar\partial\right) {\rm e}^{-\hat V}
\left(\hat\Phi - \sqrt 2 \partial\right)
{\rm e}^{\hat V}
+ \bar\partial{\rm e}^{-\hat V}\partial{\rm e}^{\hat V}
\right\}\bigg]\nonumber\\
&& + \frac{1}{4e^2}\left(\int d^2\theta\ W^\alpha W_\alpha 
+ \int d^2\bar\theta\ \bar W_{\dot\alpha} 
\bar W^{\dot\alpha}\right)\nonumber\\
&& + \frac{1}{e^2}\int d^2\theta d^2\bar\theta\ \left\{
\left(\bar\Phi - \sqrt 2 \bar \partial V\right)
\left(\Phi - \sqrt 2\partial V\right)
- \bar \partial V \partial V \right\}\nonumber\\
&& + \int d^2\theta d^2\bar\theta \left[
Q^{i\dagger} {\rm e}^{- V - \hat V} Q_i
+ \tilde Q^i {\rm e}^{V + \hat V} \tilde Q^\dagger_i\right]
\nonumber\\
&&+ \int d^2\theta\ \left[ \tilde Q^i
\left(\partial - \frac{1}{\sqrt 2}\Phi 
- \frac{1}{\sqrt 2}\hat \Phi\right) Q_i
+ \frac{v^2}{2\sqrt 2} \Phi\right]\nonumber\\
&&+ \int d^2\bar \theta\ \left[ Q^{i\dagger}
\left(-\bar\partial - \frac{1}{\sqrt 2}\bar\Phi 
- \frac{1}{\sqrt 2}\hat \Phi^\dagger\right) 
\tilde Q_i^\dagger
+ \frac{v^2}{2\sqrt 2} \bar\Phi\right]\,,
 \label{Lagrangian}
\end{eqnarray}
where contraction of $SU(N)$ flavor index $i$ 
is implied, and the normalization of the $SU(N)$ generators 
is taken as ${\rm Tr}(T^AT^B) = \delta^{AB}$. 
The 4 SUSY superfield strengths are defined by 
\begin{eqnarray}
 W_\alpha = - \frac{1}{4}\bar D\bar D D_\alpha V,\quad
\hat W_\alpha = \frac{1}{4} \bar D \bar D {\rm e}^{\hat V} 
 D_\alpha {\rm e}^{-\hat V} .
 \label{eq:super-F-st}
\end{eqnarray}
A real constant $v^2$ is called the 
Fayet-Iliopoulos parameter
\footnote{
We choose the F-type (``magnetic'') Fayet-Iliopoulos 
parameter by using $SU(2)_R$ rotation. 
}.
A holomorphic coordinate $z$ 
and the derivative $\del$ with respect to $z$ 
are used for coordinates in extra dimensions 
\begin{eqnarray}
 && z = \1{2} (x^5 + i x^6) \,, \quad 
 \partial \equiv \partial_5 - i\partial_6,\quad 
\end{eqnarray}
The Lagrangian (\ref{Lagrangian}) is invariant under 
the supergauge transformations, given by
\begin{eqnarray}
&& V \rightarrow V + \Lambda + \bar\Lambda,\quad
\Phi \rightarrow \Phi + \sqrt 2\partial \Lambda,\nonumber\\
&& {\rm e}^{\hat V} \rightarrow 
{\rm e}^{\hat \Lambda} {\rm e}^{\hat V} 
{\rm e}^{{\hat \Lambda}^\dagger},\quad
\hat \Phi \rightarrow {\rm e}^{\hat \Lambda}
\left(\hat \Phi - \sqrt 2\partial\right)
{\rm e}^{- \hat \Lambda},\\
&& Q_i \rightarrow {\rm e}^{\Lambda + \hat \Lambda}Q_i,\quad
\tilde Q^i \rightarrow \tilde Q^i 
{\rm e}^{- \Lambda - \hat \Lambda}
,\nonumber
\end{eqnarray}
with $\Lam$ ($\bar \Lam$) and 
$\hat \Lam= \sum_A \Lam^A T^A$ 
($\hat \Lam\dagg = \sum_A \bar \Lam^A T^A$) 
chiral (anti-chiral) superfields 
for $U(1)$ and $SU(N)$ gauge transformations, 
respectively. 
Note that $\Phi$ and $\hat \Phi$ also receive 
gauge transformations.

In the Lagrangian (\ref{Lagrangian}) 
the $D=6$ kinetic terms of the $U(1)$ and $SU(N)$ gauge fields 
consist of the four-dimensional part and 
the extra-dimensional part. 
The first and third lines of Eq.(\ref{Lagrangian}) 
give the four-dimensional part in terms of the ordinary 
4 SUSY superfield strengths in Eq.(\ref{eq:super-F-st}) which 
contain ordinary gauge field strengths 
\begin{eqnarray}
F_{mn} = \partial_mA_n - \partial_nA_m,\quad
\hat F_{mn} = \partial_m \hat A_n - \partial_n \hat A_m 
- \frac{i}{2}\left[\hat A_m, \hat A_n\right] .
\end{eqnarray}
On the other hand, the extra-dimensional part 
is given by the
fourth line for the $U(1)$ part 
and the second line for the $SU(N)$ part 
in the Lagrangian (\ref{Lagrangian}), respectively, 

The auxiliary fields can be eliminated by 
their algebraic equations of motion as
\begin{eqnarray}
D &=& \frac{e^2}{2}
 \left(q^{\dagger i} q_i - \tilde q^i \tilde q^\dagger{}_i\right),\quad
D^A = \frac{g^2}{2}
 \left(q^{\dagger i} T^A q_i - \tilde q^i T^A \tilde  q^\dagger{}_i\right),\\
F_\Phi &=& \frac{e^2}{\sqrt 2} 
  \left(q^{\dagger i} \tilde q^\dagger{}_i - \frac{v^2}{2} \right),\quad
F_\Phi^A = \frac{g^2}{\sqrt 2} q^{\dagger i} T^A \tilde q\dagg{}_i,\\
F_i &=& 0,\qquad
\tilde F^\dagger{}_i = 0.
\end{eqnarray}
Substituting these back into the component Lagrangian,
we obtain the scalar potential 
\begin{eqnarray}
V &=& \frac{1}{g^2}\sum_{A=1}^{N^2-1}
\left(\frac{1}{2}(D^A)^2 + |F_\Phi^A|^2\right) 
+ \frac{1}{e^2} \left( \frac{1}{2}D^2 + |F_\Phi|^2\right)
+ F^{\dagger i}F_i + \tilde F^i \tilde F^\dagger{}_i \nonumber\\
&=& g^2 \sum_{A=1}^{N^2 -1} \left[ 
\frac{1}{8}\left(q^{\dagger i} T^A q_i 
  - \tilde q^i T^A \tilde q^\dagger{}_i\right)^2 
+ \frac{1}{2}\left| q^{\dagger i} T^A 
  \tilde q^\dagger{}_i \right|^2\right]\nonumber\\
&& + e^2\left[\frac{1}{8}\left(q^{\dagger i} q_i 
   - \tilde q^i \tilde q^\dagger{}_i\right)^2 
+ \frac{1}{2}\left| q^{\dagger i} \tilde q^\dagger{}_i 
  - \frac{v^2}{2}  \right|^2\right].
\end{eqnarray}
The  supersymmetric vacuum is determined through 
the $D$-term and $F$-term flat conditions:
\begin{eqnarray}
 q^a{_i} = \tilde q\dagg{}^a{_i} = \frac{v}{\sqrt{2N}} \delta^a{_i}.
\end{eqnarray}
These vacuum expectation values (VEVs) 
break gauge symmetry to a discrete symmetry ${\bf Z}_N$ 
where broken $U(1)$ part ensures 
the topological stability of vortex configuration
\beq
 \pi_1 \left( SU(N) \times U(1) \over {\bf Z}_N \right)
 = {\bf Z} \ni k , 
 \label{eq:pi1}
\eeq
with the topological charge $k$. 
Simultaneously the VEVs break the global symmetry as 
\begin{eqnarray}
U(1)_{\rm G} \times SU(N)_{\rm G} \times SU(N)_{\rm F} 
\rightarrow
SU(N)_{\rm G+F}
\end{eqnarray}
where $U(1)_{\rm G} \times SU(N)_{\rm G}$ is 
a global transformation in the gauge symmetry. 
The unbroken group $SU(N)_{\rm G+F}$ acts on $q$ as
\beq
 q^a{}_i \to U^a{}_b q^b{}_j U^{-1}{}^j{}_i 
 \label{G+F}
\eeq
with the same group element $U$ 
in $SU(N)_{\rm G}$ and $SU(N)_{\rm F}$, 
and hence this vacuum is called the color-flavor 
locking vacuum.

\section{Non-Abelian Vortices in Six Dimensions}
\label{sc:NonAbelianVortices}

In this section we construct 
the vortex configuration which depends on $x^5,x^6$ only, 
and review solutions 
for non-Abelian vortices obtained in \cite{Hanany:2003hp} 
and \cite{Auzzi:2003fs}
in three and four dimensions, respectively. 
We will add some details of comparison between 
solutions for different values of the $U(1)$ gauge 
coupling $e$. 
The four-dimensional Lorentz invariance requires 
\begin{eqnarray}
A_m = 0,\quad \hat A_m = 0,\quad \lambda_1=0,\quad 
\lambda_2=0,\quad \psi_i=0,\quad\tilde\psi^i=0. 
\end{eqnarray}
We also demand that all the remaining dynamical fields 
$\phi,\hat\phi,q_i,\tilde q^i$ and
the auxiliary fields 
$F_\Phi,\hat F_\Phi,D,\hat D,F_i,\tilde F^i$
depend on only extra coordinates $x^5,x^6$.
Then the bosonic part of the Lagrangian reduces to 
\begin{eqnarray}
{\cal L}_{\rm boson}^{\rm (5,6)} &=& \frac{1}{g^2}{\rm Tr}
\left[ - \frac{1}{2}\left(\hat F_{56}\right)^2 
+ \frac{1}{2}\hat D^2 
+ |\hat F_\Phi|^2\right]
+ \frac{1}{e^2}\left[- \frac{1}{2}\left(F_{56}\right)^2 
 + \frac{1}{2}D^2 + |F_\Phi|^2\right]
\nonumber\\
&& - \bar {\cal D} q^{\dagger i}{\cal D}q_i 
+ \frac{1}{2}q^{\dagger i} 
\left(F_{56} + \hat F_{56}\right) q_i + F^{\dagger i}F_i
- \frac{1}{2} q^{\dagger i} \left( D + \hat D\right) q_i
\nonumber\\
&& - {\cal D} \tilde q^i \bar{\cal D} \tilde q^\dagger{}_i 
    - \frac{1}{2} \tilde q^i 
\left( F_{56} + \hat F_{56}\right) \tilde q^\dagger{}_i
+ \tilde F^i \tilde F^\dagger{}_i
+ \frac{1}{2} \tilde q^i\left(D + \hat D\right) 
  \tilde q^\dagger{}_i \nonumber\\
&&- \frac{1}{\sqrt 2} \tilde q^i 
\left( F_\Phi + \hat F_\Phi\right) q_i 
+ \frac{v^2}{2\sqrt 2} F_\Phi 
- \frac{1}{\sqrt 2} q^{\dagger i} \left(\bar F_\Phi 
+ {\hat F}_\Phi^\dagger\right) \tilde q^\dagger{}_i
+ \frac{v^2}{2\sqrt 2}\bar F_\Phi.
\end{eqnarray}
In order to obtain vortex solutions, 
we further put an Ansatz
\begin{eqnarray}
q^a{_i} 
= \tilde q\dagg{}^a{_i} \rightarrow \frac{1}{\sqrt 2}q^a{_i} 
\end{eqnarray}
reexpressing these using the same character $q^a{_i}$.
Eliminating the auxiliary fields by their algebraic 
equations of motion, 
the above Lagrangian becomes 
\begin{eqnarray}
{\cal L}_{\rm vort} &=& 
- \frac{1}{2g^2} \sum_{A=1}^{N^2-1}(F_{56}^A)^2 
- \frac{1}{2e^2}(F_{56})^2 
- \frac{1}{2}\bar{\cal D}q^{\dagger i}{\cal D}q_i 
- \frac{1}{2}{\cal D}q^{\dagger i} \bar{\cal D}q_i
- V_{\rm vort},\\
V_{\rm vort} &=& \frac{g^2}{8}\sum_{A=1}^{N^2-1}
   \left(q^{\dagger i}T^Aq_i\right)^2
+ \frac{e^2}{8}\left(q^{\dagger i}q_i - v^2\right)^2.
\end{eqnarray}
The stability of the vortices in this model is 
ensured by the nontrivial topological winding number 
(\ref{eq:pi1}) arising from the 
color-flavor locking vacuum 
\begin{eqnarray}
|q^a{_i}| = \frac{v}{\sqrt N}\delta^a{_i}.
\end{eqnarray}
The tension of a vortex solution is given by
\begin{eqnarray}
T_{\rm NA} &=&
\int d^2x\ \left[ \frac{1}{2g^2}\sum_{A=1}^{N^2-1}(F_{56}^A)^2 
+ \frac{1}{2e^2}(F_{56})^2 
+ \frac{1}{2}\bar{\cal D}q^{\dagger i}{\cal D}q_i 
+ \frac{1}{2}{\cal D}q^{\dagger i} \bar{\cal D}q_i + V\right]\nonumber\\
&=& \int d^2x\ 
\bigg[ \frac{1}{2}\sum_{A=1}^{N^2-1}
\left(\frac{1}{g}F_{56}^A - \frac{g}{2}q^{\dagger i}T^Aq_i\right)^2
+ \frac{1}{2}\left\{\frac{1}{e}F_{56} 
   - \frac{e}{2}\left(q^{\dagger i}q_i - v^2\right)\right\}^2
\nonumber\\
&&\qquad\qquad
+ \bar {\cal D} q^{\dagger i} {\cal D} q_i - \frac{v^2}{2}F_{56}\bigg],
\end{eqnarray}
with $d^2x \equiv dx^5 dx^6 = 2i d z d z^*$, where we have used 
\begin{eqnarray}
{\cal D}q^{\dagger i} \bar{\cal D}q_i 
 = \bar{\cal D}q^{\dagger i} {\cal D}q_i 
   - q^{\dagger i}\left(F_{56} + \hat F_{56} \right) q_i.
\end{eqnarray}
It is bounded from below by
\begin{eqnarray}
T_{\rm NA} \ge - \frac{v^2}{2}\int d^2x \ F_{56}.
\label{eq:NA-tension}
\end{eqnarray}
This inequality is saturated when the 
following BPS equations are 
satisfied~\cite{Auzzi:2003fs, Hanany:2003hp} 
:
\begin{eqnarray}
 F^A_{56} &=& 
 \frac{g^2}{2}q^{\dagger i}T^Aq_i,\label{NA_vortex_1}\\
 F_{56} &=& 
 \frac{e^2}{2}\left(q^{\dagger i}q_i - v^2\right),
    \label{NA_vortex_2}\\
 {\cal D}q_i &=& 0.\label{NA_vortex_3}
\end{eqnarray}
These are natural generalizations of the BPS equations 
for the 
$U(1)$ vortices~\cite{Abrikosov:1956sx,Nielsen:cs}.

The $U(1)$ vortices are solutions of 
the Abelian--Higgs model which is obtained 
from (\ref{Lagrangian}) 
by retaining only the $U(1)$ vector multiplet $(V, \Phi)$ 
and a single charged hypermultiplet $q$. 
The BPS equations in this $U(1)$ case are 
of the form: 
\begin{eqnarray}
F_{56} = \frac{e^2}{2}\left(|q|^2 - v^2\right),\quad
{\cal D}q = 0.
\label{eq:U1BPSeq}
\end{eqnarray}
If the topological winding number $\pi_1(U(1))$ 
in Eq.~(\ref{eq:pi1}) is $k$, we assume the vortex Ansatz 
with profile functions $\varphi$ 
and $f$ 
as a function of $r$ 
\begin{eqnarray}
q = {\rm e}^{-ik\theta}\varphi(r),\quad 
A_\mu = 2\varepsilon_{\mu\nu}\frac{x^\nu}{r^2}(k - f(r)),
\label{ANO_ansatz}
\end{eqnarray}
where $(r,\theta)$ are polar coordinates on the 
$x_5$--$x_6$ plane, 
$\mu,\nu$ run from 5 to 6 and 
the antisymmetric tensor is defined as $\varepsilon_{56}=1$.
Then the above BPS equations reduce to 
\begin{eqnarray}
2\frac{f'}{r} = \frac{e^2}{2}(\varphi^2 - v^2),\quad
r\varphi' = f\varphi.\label{ANO}
\end{eqnarray}
The profile functions have to satisfy the following 
boundary conditions 
\begin{eqnarray}
f(\infty)=0,\quad \varphi(\infty) = v
,\quad
f(0) = k
,
\end{eqnarray}
\begin{eqnarray}
\varphi(0)=0
,\quad 
{\rm if} \quad k \not=0. 
\end{eqnarray}
The tension of the $U(1)$ BPS vortex with the winding 
number $k$ is given by~\cite{Abrikosov:1956sx,Nielsen:cs} 
\begin{eqnarray}
T_{U(1)
} = - \frac{v^2}{2}\int d^2x\ F_{56} 
= - \frac{v^2}{2}\int d^2x\ 2 \frac{f'}{r} 
= - 2\pi v^2\big[ f \big]^\infty_0 = 2\pi k v^2.
\label{eq:U1tension}
\end{eqnarray}
We will call the $U(1)$ vortex with the unit winding number 
as ANO vortex whose tension is given by 
\begin{eqnarray}
T_{\rm ANO}=2\pi v^2. 
\label{eq:ANO-tension}
\end{eqnarray}
The existence of the solution 
for the $U(1)$ BPS equations (\ref{ANO}) 
with the winding 
number $k$ 
has been demonstrated and its power series 
expansion has been studied\cite{dVegaSchaposnik}, 
although no explicit analytic solution has been obtained 
so far. 

Let us turn our attention to the non-Abelian vortices 
by solving 
Eqs.~(\ref{NA_vortex_1})--(\ref{NA_vortex_3}). 
We first reexpress them 
in a matrix form 
defined by~\cite{Hanany:2003hp} 
\begin{eqnarray}
\check{A}_\mu \equiv A_\mu  {\bf 1}_N + \hat A_\mu,\quad
\check{F}_{\mu\nu} \equiv F_{\mu\nu}{\bf 1}_N + \hat F_{\mu\nu}.
\end{eqnarray}
Then Eqs.~(\ref{NA_vortex_1}) and (\ref{NA_vortex_2}) are combined into 
the following matrix equation:
\begin{eqnarray}
\check F_{56} = \frac{g^2}{2}
  \left(q_i q^{\dagger i} - \frac{v^2}{N}{\bf 1}_N\right)
+ \left(\frac{e^2}{2}- \frac{g^2}{2N}\right)
{\rm Tr}\left(q_i q^{\dagger i} - \frac{v^2}{N}{\bf 1}_N\right) 
 {\bf 1}_N .
\label{NA_vortex_matrix}
\end{eqnarray}
Let us first consider the case of general gauge couplings 
$e$ and $g$.
A natural vortex Ansatz corresponding to a highly 
symmetric point in the internal space and in the 
extra dimension, 
namely the cylindrically symmetric Ansatz 
with a common origin for all the vortices, is given by 
\begin{eqnarray}
q &=& \frac{1}{\sqrt N} \left(
\begin{array}{cccc}
{\rm e}^{-i\ell_1\theta}\varphi_1(r) & & &\\
& {\rm e}^{-i\ell_2\theta}\varphi_2(r) & &\\
& & \ddots & \\
& & & {\rm e}^{-i\ell_N\theta}\varphi_N(r)
\end{array}
\right),\\
\check A_\mu &=& 2\varepsilon_{\mu\nu}\frac{x^\nu}{r^2}
\left(
\begin{array}{cccc}
\ell_1 - f_1(r) & & &\\
& \ell_2 - f_2(r) & &\\
& & \ddots &\\
& & & \ell_N - f_N(r)
\end{array}
\right),\label{NA_ansatz}
\end{eqnarray}
where $\ell_i\ (i=1,2,\cdots,N)$ are integers. 
Inserting this Ansatz into the above matrix equation 
(\ref{NA_vortex_matrix}),
one finds~\cite{Auzzi:2003fs} 
\begin{eqnarray}
2\frac{f'_i}{r} = \frac{g^2}{2N}\left(\varphi_i^2 - v^2\right)
+ \left(\frac{e^2}{2} - \frac{g^2}{2N}\right)
\left(\frac{1}{N}\sum_{j=1}^N\varphi_j^2 - v^2\right).
\label{NA_1}
\end{eqnarray}
The remaining equation (\ref{NA_vortex_3}) gives
\begin{eqnarray}
r\varphi_i' = f_i\varphi_i.\label{NA_2}
\end{eqnarray}
Comparing these equations with Eqs.~(\ref{ANO}) for the 
$U(1)$ vortex, 
we recognize that the above Ansatz for the non-Abelian 
vortex is 
a natural extension of the Ansatz (\ref{ANO_ansatz}).
The cylindrically symmetric profile functions $(f_i,\varphi_i)$ 
should satisfy the following boundary conditions:
\begin{eqnarray}
f_i(0) = \ell_i,\quad \varphi_i(\infty) = v,\quad f_i(\infty)=0.
\label{NA_bc_1}
\end{eqnarray}
Moreover, the BPS equation (\ref{NA_2}) and 
the boundary conditions (\ref{NA_bc_1}) 
require the boundary condition for $\varphi_i(0)$
\begin{eqnarray}
\varphi_i(0) = 0,\qquad {\rm if}\quad \ell_i \neq 0.
\label{NA_bc_2}
\end{eqnarray}

Notice that Eq.~(\ref{eq:NA-tension}) implies that the 
tension of the BPS vortex is determined 
only by the $U(1)$ part of the gauge group. 
In fact, the tension of the non-Abelian vortices is given by
\begin{eqnarray}
T_{\rm NA} = - \frac{v^2}{2}\int d^2x\ F_{56}
= -\frac{v^2}{2}\int d^2x\ \frac{2}{Nr}\sum_{i=1}^Nf_i' 
= \frac{2\pi v^2}{N}\sum_{i=1}^N\ell_i.
\end{eqnarray}
This implies that 
the winding number $k \in \pi_1(U(1))$ defined in (\ref{eq:pi1}) 
is the sum of $\ell_i$:
\begin{eqnarray}
 k = \sum_{i=1}^{N} \ell_i.
\end{eqnarray}
Therefore the non-Abelian vortices with the topological winding number $k$
gives the tension which amounts to $1/N$ of the $k$ ANO vortices
\begin{eqnarray}
 T_{\rm NA} = \frac{k}{N}T_{\rm ANO}.
\end{eqnarray}
The minimal vortex solutions with 
$\ell_i =1$, $\ell_j=0 \; (j\not=i, 1\le j \le N)$ have tension 
whose value is $1/N$ of that of the ANO vortex in 
Eq.~(\ref{eq:ANO-tension}). 

When we have no particular relation between 
two gauge couplings $e$ and $g$, 
Eq.~(\ref{NA_1}) shows that the profile functions $f_i$ 
and $\varphi_i$ satisfy coupled differential equations 
whose solutions 
are generally 
different from those for the $U(1)$ 
vortices. 
Solutions 
have been obtained numerically in the case of 
$N=2$ 
in Ref.~\cite{Auzzi:2003fs}. 
They give both profile functions 
$\varphi_1$ and $\varphi_2$ which are nontrivial 
and are different from the $U(1)$ vortex 
solutions\cite{Auzzi:2003fs}. 

In the case of $\ell_1 = \ell_2 = \cdots = \ell_N 
= \ell 
$, however, 
the cylindrically symmetric boundary conditions 
(\ref{NA_bc_1}) and (\ref{NA_bc_2}), 
and the BPS equations (\ref{NA_1}) and (\ref{NA_2}) 
allow a solution for 
the profile functions which become identical 
to the vortex solutions for the $U(1)$ BPS equations 
(\ref{ANO}). 
To see this, we note that the boundary conditions 
(\ref{NA_bc_1}) and (\ref{NA_bc_2}) in this 
case allow the following Ansatz 
\begin{eqnarray}
\varphi_i(r) = \varphi(r),\quad
f_i(r) = f(r),\quad(i=1,2,\cdots,N),
\end{eqnarray}
in Eqs.~(\ref{NA_1}) and (\ref{NA_2}).
Then we find that the BPS equations for 
$f$ and $\varphi$ become completely identical to those of 
the $U(1)$ vortex~\cite{Auzzi:2003fs}:
\begin{eqnarray}
2 \frac{f'}{r} = \frac{e^2}{2}\left(\varphi^2 - v^2\right),
\quad
r\varphi' = f\varphi.
\end{eqnarray}
In contrast to the minimal vortices, 
this configuration with 
 $\ell_1 = \ell_2 = \cdots = \ell_N 
= \ell 
$ winds $\ell$ full turns around 
the $U(1)$ part without going through the $SU(N)$ part. 
Notice that the solution is valid for generic values of 
$U(1)$ gauge coupling $e$ and that 
the tension of this 
 $\ell_1 = \ell_2 = \cdots = \ell_N 
= \ell $ vortex 
is the same as that for the $\ell$ ANO vortices as shown in 
Eqs.~(\ref{eq:U1tension}) and (\ref{eq:ANO-tension}). 

Next let us consider the special case where
the gauge couplings $g$ and $e$ satisfy 
a particular relation~\cite{Hanany:2003hp} 
\beq
 g^2 = N e^2 \;. \label{ge}
\eeq
In this case the second term of the right hand side of 
Eq.~(\ref{NA_vortex_matrix}) vanishes.
This implies that the diagonal entries have no interaction 
with each other, if we stick to the configurations with 
vanishing off-diagonal entries. 
We denote 
diagonal entries of 
such a configuration as 
$q_{\star}^{(i)}$ and 
$F_{56\star}^{(i)}$ 
($i = 1,\cdots,k$) 
\beq
  (\check F_{56})^a{_b} = 
 \begin{pmatrix}
   F_{56\star}^{(1)} & & & \\
   & F_{56\star}^{(2)} & & \\
   & & \ddots &  \\
   & & & F_{56\star}^{(n)} \\
 \end{pmatrix} , \;\; 
 q^a{_i} = \frac{1}{\sqrt N}
 \begin{pmatrix}
   q_{\star}^{(1)} & & & \\
   & q_{\star}^{(2)} & & \\
   & & \ddots & \\
   & & & q_{\star}^{(n)} \\
 \end{pmatrix} .
\eeq
Eqs.~(\ref{NA_vortex_3}) and (\ref{NA_vortex_matrix}) 
give the $N$ decoupled sets of the 
BPS equations for 
this configuration 
\begin{eqnarray}
 F_{56\star}^{(i)} 
= \frac{e^2}{2} \left(|q_\star^{(i)}|^2 - v^2\right),\quad
 {\cal D}q_\star^{(i)} = 0,  
 \qquad (i=1, \cdots, N).
\end{eqnarray}
which turn out to be identical to the BPS eqs.~(\ref{eq:U1BPSeq}) 
for the $U(1)$ theory. 
The solution for each 
component is given by 
the following vortex Ansatz 
\begin{eqnarray}
q_\star^{(i)} 
= {\rm e}^{-i\ell_i\theta_i}\varphi_i(r_i),\quad
A_{\star\mu}^{(i)} 
= 2\varepsilon_{\mu\nu}\frac{x^\nu}{r_i^2}
(\ell_i - f_i(r_i)),
\end{eqnarray}
where $(r_i,\theta_i)$ are polar coordinates on the 
$x_5$--$x_6$ plane whose 
origins are generally different from each other.
Thus we obtain $N$ sets of BPS equations for the $U(1)$ 
vortices: 
\begin{eqnarray}
2\frac{f_i'}{r_i} 
= \frac{e^2}{2}\left(\varphi_i^2 - v^2\right),\quad
r_i\varphi_i' 
= f_i\varphi_i,\quad (i=1,2,\cdots,N).
\end{eqnarray}
We should stress that the cylindrical symmetry with the common 
origin for all $i=1, \cdots, N$ is not needed, since diagonal 
entries are decoupled unlike the 
case of general gauge couplings $e$ and $g$. 
The boundary conditions are given by 
\begin{eqnarray}
f_i\big|_{r_i=0} = \ell_i,\quad
\varphi_i\big|_{r_i=0} = 0,\qquad
f_i\big|_{r_i=\infty} = 0,\quad
\varphi_i\big|_{r_i=\infty} = v.
\end{eqnarray}
In this case the minimal solution with $\ell_i = 1$, 
$\ell_j = 0, j\not=i\; (1\le j \le N)$ 
is simply given by 
$\varphi_{j\neq i} = v$ and $f_{j\neq i}=0$, and 
the nontrivial profile functions $\varphi_i$ 
and $f_i$ are exactly
the same as those for the ANO vortex. 
Taking the case of $\ell_1= 1$ as an example, the Ansatz 
for a minimal vortex reduces to
\begin{eqnarray}
q = \frac{1}{\sqrt N}\left(
\begin{array}{cccc}
{\rm e}^{-i\theta}\varphi_1 & & & \\
 & v & &\\
& & \ddots &\\
& & & v
\end{array}
\right),\ 
\check A_\mu = 2\varepsilon_{\mu\nu}\frac{x^\nu}{r^2}\left(
\begin{array}{cccc}
1 - f_1 & & & \\
 & 0 & &\\
& & \ddots &\\
& & & 0
\end{array}
\right).\label{NA_single}
\end{eqnarray} 
This type of solutions has been considered in 
Ref.~\cite{Hanany:2003hp}
for the case of three spacetime dimensions.

\section{Effective Field Theory on Vortices}
\label{sc:EffectiveTheory}

In this section we discuss the effective field 
theory on vortices restricting 
to the case of $g^2 = N e^2$ in Eq.~(\ref{ge}). 
When Eq.~(\ref{ge}) does not hold, 
the effective theory is expected to be deformed. 
We will return to this problem in Sec.~5. 

In the first subsection, we discuss the effective field theory 
on non-Abelian vortices in a purely field 
theoretical context, by starting from a single vortex or 
noninteracting multi-vortex solutions. 
It works well for a single vortex but 
not for multi-vortices because 
we do not find correct number of bosons 
needed for the moduli space dimension.
In the second subsection, 
we discuss the moduli space for 
multi-vortices inspired by branes which are 
proposed by Hanany and Tong
\cite{Hanany:2003hp}, 
and find that their effective theory contains 
a number of extra massless bosons, which are called 
quasi-Nambu-Goldstone bosons.  
Combined with the genuine Nambu-Goldstone bosons, 
we obtain all the massless bosons, 
exhausting the correct dimensions of the moduli space.
As shown in \cite{Hanany:2003hp}, 
the Hanany-Tong metric does not coincide with 
the Manton metric even for two vortices in the $U(1)$ gauge theory. 
Hence we do not expect that it is a correct metric 
describing  the scattering of vortices 
but that it has correct dimensions, topology and symmetry. 
We discuss this problem in the next section. 
Moreover the brane construction in \cite{Hanany:2003hp} was 
given for three-dimensional theory and later 
it was generalized to four dimensions 
in \cite{HT2} taking a T-dual. 
We simply use the same Lagrangian 
for six dimensions because the structure of the moduli space 
is independent of dimensions at least classically.

\subsection{Orientation Moduli for Vortices}
Dynamics of vortices is described by the effective field 
theory of massless fields on the vortices. 
The Manton method~\cite{Ma} or the 
mode expansion method can be used to 
construct the effective field theory, 
which is in general a nonlinear sigma model 
with the moduli space of vortices as its target manifold.
Since we do not have explicit solutions even 
for the ANO vortex of the $U(1)$ gauge theory, 
it is difficult to integrate over the extra dimensions. 
However in the case of a single non-Abelian vortex, 
the moduli space metric is determined by symmetry only 
with no need of the explicit solution. 

First, let us consider 
the effective field theory on 
a single non-Abelian vortex.
The BPS equations (\ref{NA_vortex_1})--(\ref{NA_vortex_3}) 
are invariant under a global symmetry $SU(N)_{\rm G+F}$ 
given in (\ref{G+F}). 
Namely, when a set of $\check F_{56}$ and $q$ 
is a solution, then the following
transformation keeps it as a solution of the BPS equations
\begin{eqnarray}
\check F_{56} \rightarrow U_{\rm G}\ \check F_{56}\ U_{\rm G}^{-1},\quad
q \rightarrow U_{\rm G}\ q\ U_{\rm F}\quad
\left(U_{\rm G} \in SU(N)_{\rm G},\ U_{\rm F} = U_{\rm G}^{-1} \in SU(N)_{\rm F}\right).
\end{eqnarray}
Since the configuration at 
the spatial infinity in the $x^5$-$x^6$ plane 
is fixed under this transformation, the 
corresponding zero modes should be normalizable 
physical modes. 
This fact implies that the moduli space metric admits 
an isometry $SU(N)_{\rm G+F}$. 
On the other hand, our Ansatz (\ref{NA_single}) is 
invariant under the subgroup 
$H \equiv SU(N-1) \times U(1)$ 
of $SU(N)_{\rm G+F}$. 
So $SU(N)_{\rm G+F}$ is {\it spontaneously} broken down 
to its subgroup $H$ by the 
vortex configuration (\ref{NA_single}), 
and $H$ is the isotropy group of the moduli space metric.
We thus have observed that the moduli space 
for a single non-Abelian vortex, denoted by ${\cal M}_{1,N}$, 
can be written as~\cite{Hanany:2003hp,Auzzi:2003fs}
\begin{eqnarray}
{\cal M}_{1,N} = {\bf C} \times 
\frac{SU(N)_{\rm G+F}}{SU(N-1) \times U(1)}
\simeq {\bf C} \times {\bf C}P^{N-1} \, .
\end{eqnarray}
The first factor ${\bf C}$ corresponds 
to the Nambu-Goldstone degree of freedom for
the vortex position $z_0$ in the $x^5$-$x^6$ plane 
which exists for Abelian cases also. 
Whereas the second part is parametrized by 
Nambu-Goldstone bosons arising from 
spontaneously broken internal symmetry.\footnote{
In eight SUSY theories, domain walls are known 
to be accompanied with the 
Nambu-Goldstone bosons for broken internal $U(1)$ symmetry~\cite{DW}.
} 
This is a characteristic feature in non-Abelian cases.
The complex dimension of the above metric is $N$ 
in agreement with the index theorem 
calculated by Hanany and Tong 
\cite{Hanany:2003hp} generalizing the Abelian case \cite{Weinberg},  
and therefore we do not have any other massless modes.

The moduli space of a single vortex has been 
completely determined by symmetry only. 
In the case of  multi-vortices, we need more 
considerations as shown below.
Let us denote the moduli space of $k$ vortices 
by ${\cal M}_{k,N}$ and consider the case of $k \leq N$:
\beq
  (\check F_{56})^a{_b} = 
 \begin{pmatrix}
   F_{56\star}^{(1)} & & & & &  \\
   & \ddots & & & & \\
   & & F_{56\star}^{(k)} & & &  \\
   & & & 0 & &  \\
   & & & & \ddots & \\
   & & & & &  0 \\
 \end{pmatrix} , \;\; 
 q^a{_i} = \frac{1}{\sqrt N}
 \begin{pmatrix}
   q_{\star}^{(1)} & & & & &  \\
   & \ddots & & & & \\
   & & q_{\star}^{(k)} & & &  \\
   & & & v & &  \\
   & & & & \ddots & \\
   & & & & &  v \\
 \end{pmatrix}. \label{k-ansatz}
\eeq
Here,
$q_{\star}^{(I)}$ and $F_{56\star}^{(I)}$ 
($I = 1,\cdots,k$) are solutions 
of the decoupled $U(1)$ vortices : 
\begin{eqnarray}
q_{\star}^{(I)} = q_{\star}(z-z_I,z^*-z_I^*), 
\quad  
F_{56\star}^{(I)} = F_{56\star}(z-z_I,z^*-z_I^*), 
\qquad 
I = 1,\cdots,k, 
\end{eqnarray}
where $z_I$ are the positions of the $I$-th vortex. 
Since 
these $k$ vortices 
do not interact with each other in this particular Ansatz 
of diagonal embedding, which corresponds to a 
particular sector (submanifold) of the full moduli space.
Let us note that the above solution of non-Abelian vortices 
is exact irrespective of the position of the vortices 
$z_I$. 
Therefore these moduli $z_I$ are exact 
translational moduli of the $k$-vortex solution. 

The symmetry $SU(N)_{\rm G+F}$ is spontaneously 
broken by the configuration (\ref{k-ansatz}) down into 
$SU(N-k) \times U(1)^k$ provided 
$z_I \neq z_J$ for $I \neq J$. 
Hence the Nambu-Goldstone bosons for this breaking appear 
and parametrize the coset 
$SU(N)/[SU(N-k) \times U(1)^k]$ 
inside the moduli space of $k$-vortices 
${\cal M}_{k,N}$.
We also have moduli $z_I$ for translation of vortices. 
Therefore 
the moduli space certainly contains 
\beq
{\bf C}^k \times 
{SU(N)_{\rm G+F} \over SU(N-k) \times U(1)^k} 
  \;\;\; (\subset {\cal M}_{k,N} ) \,,
  \label{NG-coset1}
\eeq
as a submanifold. 
This does not coincides with the whole 
moduli space of $k$-vortices since the complex 
dimension of (\ref{NG-coset1}) is  $kN - \1{2} k (k-1)$, 
which is less than 
$\dim_{\bf C} {\cal M}_{k,N} = kN$ 
calculated using the index theorem~\cite{Hanany:2003hp}.
This comes from the fact that we are considering a 
particular Ansatz (\ref{k-ansatz}) with too small number of 
parameters. 
We thus have missed $\frac{k (k-1)}{2}$ additional complex moduli 
parameters, which will turn 
out to be the so-called quasi-Nambu-Goldstone modes 
when we consider the low-energy effective 
Lagrangian~\cite{BLPY}--\cite{HNOO} 
in the next subsection.

In the case $z_I = z_J$ for 
a set of $I,J$ with $I\neq J$ 
the unbroken symmetry is enhanced to 
$SU(N-k) \times U(2) \times U(1)^{k-2}$, 
and we find the moduli space includes 
\beq
 {\bf C}^k \times {SU(N)_{\rm G+F} \over 
  SU(N-k) \times U(2) \times U(1)^{k-2}} 
  \;\;\; (\subset {\cal M}_{k,N} )
   \label{NG-coset2}
\eeq
as a submanifold with the complex dimension 
$kN - \1{2}k(k-1) -1$. 
Similarly, depending on how many translational 
moduli $z_I$ coincide, 
several different coset manifolds of Nambu-Goldstone 
modes are 
embedded into the moduli space. 
Let us consider the case of $n_a$ 
($a=1,\cdots,m$; $\sum_{a=1}^m n_a \leq k$) of them 
coincide like
\beq
\vspace{-1cm}   (z_1,\cdots,z_k) 
 = (\underbrace{z_{(1)},\cdots,z_{(1)}}_{n_1};
\cdots ;\underbrace{z_{(m)},\cdots,z_{(m)}}_{n_m};
z_{(m+1)};z_{(m+2)};\cdots;z_{(k - \sum_{a=1}^m n_a + m)}).
 \label{zs}
\eeq
Then the coset manifolds of Nambu-Goldstone 
modes 
\beq
  {\bf C}^k 
   \times {SU(N)_{\rm G+F} \over SU(N-k) \times \prod_{a} 
   SU(n_a) \times U(1)^{k + m - \sum_a n_a}} \;
    \;\;\; (\subset {\cal M}_{k,N} ) \,,
     \label{NG-coset3}
\eeq
with the complex dimension 
$Nk - \1{2}k(k-1) - \sum_{a=1}^m \1{2} n_a (n_a-1)$,
is embedded into the full moduli space. 

In either case we did not find enough number of 
massless bosons parametrizing the full moduli space.
Instead we have several coset spaces with various dimensions 
parametrized by 
Nambu-Goldstone bosons for several breaking patterns 
of internal symmetry $SU(N)_{\rm G+F}$.
This is actually a feature for symmetry breaking  
in SUSY (gauge) theories with quasi-Nambu-Goldstone bosons~\cite{KS,Ni}.
As discussed in the next subsection,
such a situation can be completely described by 
the moduli space theory formulated by a SUSY gauge theory 
(or a K\"ahler quotient) 
which was proposed by Hanany and Tong 
(using the brane configuration in 
the string theory)~\cite{Hanany:2003hp}.

Before closing this subsection, we make a comment. 
When we embed the solutions of the 
$U(1)$ vortices into diagonal entries in 
$(\check F_{56})^a{_b}$ and 
$q^a{_i}$  in Eq.~(\ref{k-ansatz}), 
these vortices do not interact. 
However, if we would like to consider an Ansatz 
of vortices placed in off-diagonal entries unlike 
Eq.~(\ref{k-ansatz}), 
it is  no longer an exact solution of the BPS equations. 
This fact implies that these vortices are interacting. 
These vortices become exact solutions only in the limit 
of infinite separation between vortices. 
We cannot avoid this situation in the case of large numbers of 
vortices $k \geq N$.

\subsection{Symmetry Structure in the Vortex Moduli Space}
\label{sc:sym-moduli-space}
Inspired by brane configurations, an effective theory 
on $k$-vortices in the 
$U(N)$ gauge theory with $N$ hypermultiplets was 
given by Hanany and Tong~\cite{Hanany:2003hp} 
in the form of a K\"ahler quotient. 
It is described by $D=4$, ${\cal N}=1$ SUSY 
(with four supercharges) $U(k)$ gauge theory 
with {\it auxiliary} vector superfields 
of a $k$ by $k$ matrix $V(x^m,\theta,\thb)$ 
(without kinetic term)
and  matter chiral superfields of 
a $k$ by $N$ matrix $\ph(x^m,\theta,\thb)$ and 
a $k$ by $k$ adjoint matrix $Z(x^m,\theta,\thb)$. 
Their $U(k)$ supergauge transformation laws are given by
\beq
 && e^V \to e^{V'} = e^{-\Lam\dagg} e^V e^{-\Lam} \,,\non
 && Z \to Z' = e^{\Lam} Z e^{-\Lam} \,,\non
 && \ph \to \ph' = e^{\Lam} \ph \,, \; \label{gauge}
\eeq
with $\Lam (x^m,\theta,\thb)$ a $k$ by $k$ matrix 
chiral superfield for a gauge parameter. 
Note that the gauge symmetry is 
complexified to $U(k)^{\bf C} = GL(k,{\bf C})$.
The effective Lagrangian proposed by Hanany and Tong is  
\beq
 {\cal L}_{\rm HT} = \int d^4 \theta 
  \left[ \tr (\ph\ph\dagg e^V) + \tr (Z\dagg e^V Z e^{-V}) 
   - c \, \tr V \right] \label{HT}
\eeq
with $c$ a Fayet-Iliopoulos parameter 
related to the gauge coupling $g$ of the 
original gauge theory by $c = 2\pi/g^2$. 
This theory has a global symmetry $SU(N)$ 
originated from $SU(N)_{\rm G+F}$:
\beq
 \ph \to \ph' = \ph U \;, \label{global}
\eeq
with $U$ the $SU(N)$ unitary matrix in the fundamental 
representation.
The Higgs branch of this theory is an $U(k)$ K\"ahler quotient 
${\cal M}_{k,N} = \{Z,\ph\}//U(k) = \{Z,\ph\}/U(k)^{\bf C}$ 
with $\dim_{\bf C} {\cal M}_{k,N} = kN$,
and the theory is described by 
a nonlinear sigma model with ${\cal M}_{k,N}$ as 
the target space.
Since we have chosen the minimal kinetic term as a first guess, 
we do not expect that the metric ${\cal M}_{k,N}$ coincides 
exactly with the Manton metric, 
although their dimension, topology and symmetry should coincide.

Here we investigate the geometric structure of the 
moduli space. 
The Wess-Zumino gauge makes physical content transparent, but 
breaks  $U(k)^{\bf C}$ to $U(k)$. 
In the following we do not take the Wess-Zumino gauge, and 
use the full complex gauge symmetry for 
fixing superfields. 
First, in the $k=1$ case of the single vortex, 
the Lagrangian (\ref{HT}) is the K\"ahler quotient 
formulation for ${\bf C}P^{N-1}$~\cite{DDL,HN1}. 
Eliminating $V$ by its equation of motion, 
$\ph \ph\dagg e^V - c =0$, 
we obtain the K\"ahler potential for 
the Fubini-Study metric on ${\bf C}P^{N-1}$, 
$K = c \log (1 + \hat\ph \hat\ph\dagg)$ 
with a gauge fixing $\ph = (1,\hat \ph)$.
In this case the moduli space agrees with 
the one parametrized by the Nambu-Goldstone bosons only,  
as discussed in the last subsection. 

Next we investigate the moduli space geometry in the 
case of $1<k \leq N$. 
In this case, it is difficult to eliminate $V$ 
in the superfield formalism.
Instead, we fix the gauge without eliminating $V$. 
Although we cannot get the K\"ahler potential in terms of 
independent fields we can identify fields as 
Nambu-Goldstone or quasi-Nambu-Goldstone bosons.
We will see that there exist some 
quasi-Nambu-Goldstone bosons which are not related to 
the spontaneously broken global symmetry. 
First of all, using the gauge symmetry (\ref{gauge}), 
$Z$ can be diagonalized as
\beq
 Z_{\rm fixed} 
  = \begin{pmatrix}
           z_1 &        & {\bf 0}\cr 
               & \ddots & \cr
       {\bf 0} &        & z_k
      \end{pmatrix} \; \label{VEV1}
\eeq
where each $z_I$ ($I=1,\cdots,k$) 
is interpreted as the position of each vortex.
The complex gauge group $U(k)^{\bf C}$ is broken down to 
its subgroup $\{U(1)^{\bf C}\}^k$ 
by the VEVs of (\ref{VEV1}), 
provided $z_I \neq z_J$ for $I\neq J$.
Using this unbroken gauge symmetry, $k$ components in 
$\ph$ can be fixed to unity (we take them to be diagonal 
entries) :
\beq
 \ph_{\rm fixed}  
 = \left(\begin{array}{ccccc|c}
          1 & D_{11} & \cdots & \cdots & D_{1,k-1}      &         \\
      C_{11}& 1 & \ddots &        & D_{2,k-1} &         \\ 
\vdots &\ddots & \ddots & \ddots & \vdots 
& {E}_{k \times (N-k)}\\ 
C_{1,k-2} & \cdots    & \ddots & 1  & D_{k-1,k-1}&   \\
  C_{1,k-1} & C_{2,k-1} & \cdots & C_{k-1,k-1} & 1    &   \\
   \end{array} \right)  \label{ph}
\eeq
with $C_{IJ}$ ($I \leq J =1,\cdots,k-1$),  
$D_{IJ}$ ($I \geq J =1,\cdots,k-1$) and $E$ 
chiral superfields. 
Note that $\ph$ has no independent degrees of freedom 
in the Abelian case $N=1$; 
it parametrizes the orientation of the vortices
in the internal space in the non-Abelian case. 

After fixing the gauge, we take the VEVs of 
these dynamical fields. 
Fluctuations of these fields from those VEVs are 
identified with propagating fields. 
In our case they are interpreted as 
Nambu-Goldstone or quasi-Nambu-Goldstone bosons. 
Depending on VEVs, 
the numbers of these bosons can vary 
with their total number unchanged~\cite{KS,Ni}.
VEVs transformed by the global symmetry 
(\ref{global}) 
are equivalent to each other, 
and so physics for instance the identification of 
the Nambu-Goldstone or quasi-Nambu-Goldstone bosons is unchanged.
If we compare two VEVs which are not related by 
the global symmetry, physics generically changes:  
unbroken symmetry can change, 
the number of Nambu-Goldstone and quasi-Nambu-Goldstone can vary 
or, even in the case of the same unbroken symmetry, 
decay constants of Nambu-Goldstone bosons are different in general.

Therefore when we take the VEVs of 
(\ref{VEV1}) and (\ref{ph}),  
to identify Nambu-Goldstone or quasi-Nambu-Goldstone bosons, 
we can use the global $SU(N)$ transformations (\ref{global}). 
Since these  $SU(N)$ transformations generically break 
the gauge fixing conditions (\ref{ph}) 
for $\{U(1)^{\bf C} \}^k$, 
we need to perform supplementary $\{U(1)^{\bf C} \}^k$ 
transformations to pull back to the gauge fixing 
hypersurface (\ref{ph}). 
The (representative of) VEVs of $\ph$ are given by 
\beq
 \left< \ph \right> 
 = \left(\begin{array}{ccccc|c}
      1 & 0 & \cdots & \cdots & 0      &         \\
      * & 1 & \ddots &        & \vdots &         \\ 
 \vdots &\ddots&\ddots& \ddots & \vdots & 
 {\bf 0}_{k \times (N-k)}\\ 
      * &\cdots&\ddots& 1      & 0      &         \\
      * & * & \cdots & * & 1   &         \\
   \end{array} \right) \label{VEV2}.
\eeq 
$\1{2} k(k-1)$ lower left components denoted by $*$ 
are not fixed by any symmetry 
and so are identified with the VEVs 
of quasi-Nambu-Goldstone bosons. 
When we change these parameters, the homogeneous space 
representing the Nambu-Goldstone bosons is gradually deformed. 
When there are generically $n$ parameters to deform a 
homogeneous space, it is called cohomogeneity $k$. 
By counting $*$'s in (\ref{VEV2}) and $z_I$ in (\ref{VEV1}) 
we find that the moduli manifold 
is of cohomogeneity $2 [\1{2}k(k-1)+k] = k(k+1)$. 
Therefore $C_{IJ}$ in the decomposition 
(\ref{ph}) can be identified as the quasi-Nambu-Goldstone 
bosons, which are the parameters of the solution without 
accompanying spontaneously broken global symmetries. 

Let us consider the case where all the $*$'s in 
(\ref{VEV2}) vanish. 
Then the total symmetry $U(k)_{\rm g} \times SU(N)_{\rm f}$ 
is broken down to a global symmetry 
$SU(N-k)_{\rm f} \times U(1)_{\rm g + f} ^k$ 
by VEVs (\ref{VEV1}) and (\ref{VEV2}) 
with no unbroken gauge symmetry 
(the suffixes g and f represent the gauge and flavor 
symmetries in the vortex world-volume theory (\ref{HT})).
Here $U(1)_{\rm g + f}^k$ is given by 
the $I$-th rotation ($I=1, \cdots,k$ 
of unbroken $U(1)^k$ gauge symmetry 
and an $SU(N)$ generator 
\beq
{\rm diag. } (0, \cdots, 0,\underbrace{N-k}_{\mbox{$I$-th}}, 
0, \cdots, 0, \underbrace{- 1,\cdots,-1}_{N-k}) \nonumber
\eeq
with the opposite angles. 
By this breaking 
there appear Nambu-Goldstone bosons parametrizing
$SU(N)/SU(N-k) \times U(1)^k$. 
Comparing (\ref{ph}) and (\ref{VEV2}) 
these Nambu-Goldstone bosons sit in $D$ and $E$ in (\ref{ph}).
The quasi-Nambu-Goldstone bosons sit in 
real $2k$ components $z_I$ in Eq.~(\ref{VEV1}) and 
$k(k-1)$ components $C_{IJ}$ in Eq.~(\ref{ph}). 
They parametrize non-compact directions of 
the moduli space ${\cal M}_{k,N}$ 
while the Nambu-Goldstone bosons parametrize compact directions. 
Therefore ${\cal M}_{k,N}$ can be locally written as\footnote{
Note that the expression by the direct product 
is not rigorous in the mathematical sense 
because we decomposed it in the level of the algebra 
but not of the group.
}
\beq
 {\cal M}_{k,N} = 
  {\bf C}^k \times {\bf C}^{\1{2}k(k-1)} 
   \times {SU(N)_{\rm G+F} \over SU(N-k) \times U(1)^k}. \; 
    \label{decom}
\eeq 
One should note that the index G+F represents the diagonal 
symmetry group for the original theory 
instead of the vortex theory. 
Comparing this with (\ref{NG-coset1}) we have found 
additional bosons with correct total number of bosons.

In (\ref{decom}) the quasi-Nambu-Goldstone bosons 
in the $U(k)$ gauge theory are 
decomposed into first two elements ${\bf C}^k$ 
and ${\bf C}^{\1{2}k(k-1)}$. 
The first one ${\bf C}^k$ corresponds to 
(approximate) Nambu-Goldstone bosons for 
translational symmetry broken 
by a configuration of vortices. 
One of them ${\bf C}$ is an exact Nambu-Goldstone boson 
corresponding to the center of mass, 
if we consider the spacetime symmetry also. 
The moduli metric for this Nambu-Goldstone boson 
is flat and forms a direct product with other moduli. 
The translation of each individual vortex around 
the center of mass becomes 
a symmetry only asymptotically in the limit of 
large separations between vortices. 
It is a parameter of the solution (moduli), 
since the translational symmetry is 
valid except finite regions of relative distances 
and it is spontaneously broken. 
These relative translation bosons ${\bf C}^{k-1}$ 
are not exact Nambu-Goldstone bosons, 
and are called a quasi-Nambu-Goldstone boson. 
The moduli metric for these bosons ${\bf C}^{k-1}$ 
is not a direct product with others, but 
sit in a fiber as we will soon see. 
The second factor ${\bf C}^{\1{2}k(k-1)}$ 
in Eq.(\ref{decom}) corresponds to 
quasi-Nambu-Goldstone bosons which are accompanied 
by the Nambu-Goldstone bosons 
$SU(N)_{\rm G+F} / [SU(N-k) \times U(1)^k]$ 
for internal symmetry broken by orientation of vortices 
in the internal space. 
They are required by the unbroken SUSY and 
were missed in the purely field theoretical  
consideration (\ref{NG-coset1}) in the last subsection.

An advantage of the vortex theory (\ref{HT}) is that 
it can describe the configurations of 
coincident or nearly coincident vortices 
without any particular Ansatz. 
Therefore we can cover values of the orientational moduli 
(quasi-Nambu-Goldstone bosons) 
which are not covered by the Ansatz (\ref{k-ansatz}). 
Let us consider a generic situation where some of 
vortices are coincident at several points $z_I$. 
If some of $z_I$ coincide, 
the unbroken symmetry is enhanced 
and the number of Nambu-Goldstone bosons is reduced. 
Since the total number of massless bosons is unchanged, 
the number of quasi-Nambu-Goldstone bosons increases 
accordingly. 
Let $n_a$ ($a=1,\cdots,m$; $\sum_{a=1}^m n_a \leq k$) 
of them coincide as in (\ref{zs})
\beq
 Z = {\rm diag. } (\underbrace{z_{(1)},\cdots,z_{(1)}}_{n_1};
\cdots ;\underbrace{z_{(m)},\cdots,z_{(m)}}_{n_m};
z_{(m+1)};z_{(m+2)};\cdots;z_{(k - \sum_{a=1}^m n_a + m)}).
\eeq
These VEVs break the gauge symmetry $U(k)$ down 
to $\prod_{a=1}^m U(n_a) \times U(1)^{k - \sum n_a}$. 
By using similar argument as above, 
the moduli manifold looks like
\begin{equation}
 {\cal M}_{k,N} = 
  {\bf C}^k \times {\bf C}^{\1{2}k(k-1) 
                 + \sum_{a=1}^m \1{2} n_a (n_a-1)} 
   \times {SU(N)_{\rm G+F} \over SU(N-k) \times \prod_{a} 
   SU(n_a) \times U(1)^{k + m  - \sum_a n_a}}.
    \label{decom2}
\end{equation}
We succeeded to obtain the second factor 
${\bf C}^{\1{2}k(k-1) + \sum_{a=1}^m \1{2} n_a (n_a-1)}$ 
of quasi-Nambu-Goldstone bosons which 
was missed in (\ref{NG-coset3}) in 
the field theoretical approach in the 
last subsection. 
Compared to the coincident vortices in Eq.(\ref{decom}), 
the unbroken symmetry is enhanced and 
Nambu-Goldstone bosons decreases whereas 
the total dimension of moduli space is unchanged. 
Hence some of the Nambu-Goldstone bosons in the case of 
non-coincident vortices 
become the quasi-Nambu-Goldstone bosons in the 
case of (partially) coincident vortices. 
(This corresponds to singular points in ${\cal M}_{k,N}/G$ 
with $G=SU(N)$~\cite{Ni}.)

The most symmetric point occurs if all of 
$Z_i$ coincide giving $Z=z_0 {\bf 1}_k$. 
At this point the formula (\ref{decom2}) 
for the moduli space gives 
\beq
 {\cal M}_{k,N} = 
  {\bf C}^k \times {\bf C}^{k(k-1)} \times G_{N,k} \; 
   \label{Gr}
\eeq
with $G_{N,k}$ the complex Grassmann manifold 
parametrizing the Nambu-Goldstone bosons for 
internal symmetry (orientational zero modes) 
\beq
 G_{N,k} = {SU(N)_{\rm G+F} \over 
  SU(N-k) \times SU(k) \times U(1)}\,.
\eeq
In the decompositions (\ref{decom}), (\ref{decom2}) and 
(\ref{Gr}) we only mean the direct product locally 
at a neighborhood of each point in the moduli space. 
It may not be correct globally. 
In the case of coincident vortices, however, 
the manifold is exactly a direct product of 
${\bf C}$ and a ${\bf C}^{k^2-1}$ 
bundle over $G_{N,k}$ denoted by\footnote{
We denote a fiber bundle $M$ with a fiber 
$F$ and a base $B$ by $M \simeq F \ltimes B$. 
Putting $F=0$ we get $M\simeq B$. 
}
\beq 
 {\cal M}_{k,N} = {\bf C} \times 
  ({\bf C}^{k^2-1} \ltimes G_{N,k}).
  \label{eq:coincid-moduli}
\eeq 
This can be shown as follows. 
We should obtain 
the base manifold, by eliminating the fiber. 
To see this, 
we substitute $Z = z_0 {\bf 1}_k$ into the original 
Lagrangian (\ref{HT}), and then we obtain 
\beq
 {\cal L}_{\rm reduced} = \int d^4 \theta 
  \left[ k z_0\dagg z_0 + \tr (\ph\ph\dagg e^V) 
   - c \, \tr V \right] . \label{Lag2}
\eeq
Then $V$ can be eliminated immediately using its 
algebraic equation 
of motion as $V = - \log \det \ph\ph\dagg$.
Substituting this back into (\ref{Lag2}) we obtain 
\beq
 {\cal L}_{\rm reduced} = \int d^4 \theta 
  \left[ k z_0\dagg z_0 + 
  c \log \det ({\bf 1}_k + \hat\ph\hat\ph\dagg) \right] 
\eeq
where we have fixed the complex gauge symmetry 
$U(k)^{\bf C}$ by choosing 
$\ph = ({\bf 1}_k, \hat\ph)$ with 
$\hat\ph$ a $k \times (N-k)$ matrix chiral superfield.
The second part is the well-known K\"ahler potential for 
the complex Grassmann manifold $G_{N,k}$.
This is the standard K\"ahler quotient construction 
of the Grassmann manifold~\cite{HN1}. 
Thus we obtain the metric ${\bf C} \times G_{N,k}$ 
in the limit of ${\bf C}^{k^2-1} \rightarrow 0$. 
By restoring the fiber ${\bf C}^{k^2-1}$, we obtain 
the bundle structure in Eq.(\ref{eq:coincid-moduli}).

\section{Deformation of the Vortex Moduli Metric}
\label{sc:deformation}
\subsection{Deformation}
It has been found that the K\"ahler metric on the moduli 
space obtained in the effective gauge theory approach 
of Hanany-Tong is in general different from that 
obtained in the approach of Manton, 
although they possess the same topology\cite{Hanany:2003hp}. 
This discrepancy may be explained by noting that 
different definitions of effective field are involved 
in different approaches. 
In the case of the Nambu-Goldstone bosons, 
$S$-matrix elements are unambiguously determined 
as a consequence of the spontaneously broken symmetry. 
This result is encoded in 
low-energy theorems, which determine 
the effective Lagrangians completely, 
even though possible field redefinitions may change 
the appearance of the effective Lagrangians. 
In the case of the quasi-Nambu-Goldstone bosons, 
$S$-matrix elements cannot be determined by 
spontaneously broken symmetry alone. 
The physical consequences such as the $S$-matrix 
elements explicitly depend on values of quasi-Nambu-Goldstone 
moduli parameters~\cite{KS,Ni,HNOO}. 
Therefore low-energy theorems cannot 
determine the structure of the effective Lagrangians 
even after fixing the definition of fields. 

Since different effective Lagrangians for the same system 
share the same symmetry properties, they should be related 
by deforming the Lagrangians while preserving the symmetry. 
Therefore we propose to list up all possible terms 
consistent with symmetry in order 
to construct the effective Lagrangian. 
This method should be applicable to solve ambiguities 
due to field redefinitions, even in the case of the 
Nambu-Goldstone bosons, but is particularly useful 
in the case of quasi-Nambu-Goldstone bosons. 
Different effective Lagrangians with quasi-Nambu-Goldstone 
bosons obtained in different approaches should be related 
once the relation between different parametrizations of 
quasi-Nambu-Goldstone bosons is identified. 
In our case, we can deform the metric 
preserving the $SU(N)$ isometry and SUSY. 
Such a family of metrics is expected to include 
both the Hanany-Tong metric and the Manton metric.
Moreover we have discussed the case where two 
gauge couplings of $U(1)$ and $SU(N)$ are 
related as $g^2 = N e^2$ (\ref{ge}). 
This relation is useful to obtain an exact solution 
by a simple embedding of $U(1)$ vortices. 
It is, however, desirable to be able to treat the 
case without 
this relation, letting $g$ and $e$ independent. 
In such cases, 
the vortex theory should be deformed accordingly. 
This deformation is not expected to 
alter the number of massless bosons or the topology 
of the moduli space.  
Hence the deformation that we discuss in this section 
should also describe such a case. 

For definiteness we discuss the case of two vortices $k=2$. 
In this case we can have two cases with different patterns 
of symmetry breaking, 
depending on the relative position of vortices 
\beq
 {\cal M}_{k=2,N} = 
  \begin{cases} 
  {\bf C}^2 \times {\bf C} 
   \times {SU(N) \over SU(N-2) \times U(1)^2} \;, \cr
  {\bf C}^2 \times {\bf C}^2 \times G_{N,2} \cr
  \end{cases} \;
\eeq
where the first case corresponds to separate vortices 
in Eq.(\ref{decom}) and 
the second case to coincident vortices in Eq.(\ref{Gr}). 
Since the common real dimension of factors multiplying 
the homogeneous space is six due to ${\bf C}^3$, 
the moduli space is of cohomogeneity six. 
The cohomogeneity corresponds to the number of 
zero modes, which do not originate from the genuine 
Nambu-Goldstone modes, in generic point of moduli space. 

To count the degree of freedom for the deformation, 
we construct invariants of 
$U(2)_{\rm g} \times SU(N)_{\rm f}$. 
To this end, let us define 
two by two matrices 
invariant under the flavor group $SU(N)_{\rm f}$ 
\beq
 M_1 \equiv \varphi\varphi\dagg e^V \;, \hs{10} 
 M_2 \equiv Z e^{-V} Z\dagg e^V \,,\; 
   \label{invariants}
\eeq
both of which transform as 
$M_a \to e^{\Lam} M_a e^{-\Lam}$. 
Noting the Cayley-Hamilton theorem 
$A^2 - (\tr A) A + (\det A) {\bf 1}_2 = 0$ 
for an arbitrary two by two matrix $A$, 
the independent invariants are found to be 
\beq
  I_{\alpha} \equiv \{ \tr M_1 , \tr M_2, \tr (M_1^2),  
  \tr (M_2^2), {\rm Re\,}\tr (M_1 M_2), 
   {\rm Im\,}\tr (M_1 M_2)\}
\eeq
Note that the number of invariants, six, coincides with the 
cohomogeneity of the moduli space 
${\cal M}_{k=2,N}$.
In the $U(k)$ gauge theoretical point of view, 
the most general Lagrangian is given by 
\beq
 {\cal L}_{\rm deformed}^{k=2} 
  = \int d^4 \theta \left[ f(I_{\alpha}) - c \,\tr V\right] \;
\eeq
with $f$ an arbitrary function of six arguments.

Further we have to require that 
the center of positions  
should decouple from other moduli variables 
so that ${\cal M}_{k,N} = {\bf C} \times 
\hat {\cal M}_{k,N}$ holds everywhere in the level 
of the metric.
This requires that 
$Z$ should appear quadratically. 
We thus obtain 
\beq
 {\cal L}_{\rm deformed}^{k=2}
  = \int d^4 \theta \left[ \tr M_2 + g(\tr M_1, \tr( M_1^2)) 
   - c \,\tr V\right] \; \label{deformed}
\eeq
with $g$ an arbitrary function of two arguments. 
One can expect that the interactions of the genuine 
Nambu-Goldstone boson 
corresponding to the center of mass 
translation is determined completely. 
It is perhaps surprising that the interactions of 
$k-1$ quasi-Nambu-Goldstone bosons in $Z$ 
are also determined, even though they are not constrained by 
low-energy theorems. 
Physical reason behind this fact is that 
the quasi-Nambu-Goldstone bosons become Nambu-Goldstone 
bosons asymptotically when all vortices are very 
far away from each other.  
The Hanany-Tong metric is obtained if we further choose 
the simplest case   
$g(\tr M_1, \tr M_1^2)_{\rm HT} = \tr M_1$. 
We have not yet identified explicitly the choice 
corresponding to the approach of Manton. 

For $k=1$ we have no degree of freedom for 
the deformation  
in agreement with the purely field 
theoretical argument.\footnote{
The K\"ahler potential corresponding to (\ref{deformed}) 
includes an arbitrary function $g$:
${\cal L} = \int d^4 \theta [\tr M_2 + g (M_1) - c V]$.
However arbitrariness disappears when $V$ is eliminated 
and the manifold is ${\bf C}\times {\bf C}P^N$ 
with the Fubini-Study metric~\cite{HN2}.
}
For general $k$ the Lagrangian can be written as
\beq
 {\cal L}_{\rm deformed}^{k} 
  = \int d^4 \theta \left[ \tr M_2 + 
  g(\tr M_1, \tr( M_1^2),\cdots, \tr(M_1^k)) 
   - c \,\tr V\right] \;, \label{deformed2}
\eeq
where the definitions of $M_1$ and $M_2$ are identical to 
(\ref{invariants}), 
because the Cayley-Hamilton theorem 
for an arbitrary $k$ by $k$ matrix is the form of
$A^k - (\tr A) A^{k-1} + \cdots + (\det A) {\bf 1}_k = 0$. 
The Hanany-Tong metric is still given by  
$g_{\rm HT} = \tr M_1$.

The deformation does not alter the shape of 
the compact submanifold $G/H$ in the moduli space 
parametrized by Nambu-Goldstone bosons, but 
it deforms the shape along non-compact directions 
parametrized by quasi-Nambu-Goldstone bosons 
(or radii of Nambu-Goldstone coset manifolds).

\subsection{Component Lagrangian}
We calculate the bosonic Lagrangian 
in the case of $k=2$. 
Superfields are expanded by the components fields
(in the Wess-Zumino gauge) as  
\beq
 && \varphi (x^m,\theta,\thb) = \varphi + \theta^2 F \, , \non
 && Z (x^m,\theta,\thb) = Z + \theta^2 F_Z \, , \non
 && V (x^m,\theta,\thb)= - \theta \sig^m \thb v_m 
  + \1{2} \theta^2 \thb^2 D , \non
 && e^V (x^m,\theta,\thb) = {\bf 1}_k 
 -  \theta \sig^m \thb v_m 
  + \1{2} \theta^2 \thb^2 \left(D - \1{2} v^2 \right)
\eeq
where we have written only bosonic fields. 
Each invariant can be calculated 
in terms of component fields, 
to yield
\beq
&& \tr (M_1) = \tr \varphi \varphi\dagg 
  - \theta \sig^{m} \thb \ 
   \tr (i\varphi \dellr_m \varphi\dagg + \varphi\varphi\dagg v_m) \non
&& \hs{15} 
+ \theta^2 \thb^2 \
   \tr \bigg[ FF\dagg + \1{4} \square \varphi \varphi\dagg 
      + \1{4} \varphi \square \varphi\dagg
 - \frac{1}{2} \partial_m \varphi \partial^m \varphi^\dagger  \non 
     \non
 && \hs{28} 
  - {i \over 2} (\varphi \dellr_m \varphi\dagg)v^m 
  + \1{2}\varphi\varphi\dagg \left(D - \1{2} v^2\right) \bigg] 
      \,,\non
 &&\tr (M_1^2) = \tr \left[(\varphi \varphi\dagg)^2 \right]
   - 2 \theta \sig^{m} \thb \
   \tr \left[\varphi\varphi\dagg 
     (i\varphi \dellr_m \varphi\dagg 
        + \varphi\varphi\dagg v_m) \right] \non
 && \hs{20} + \theta^2 \thb^2 
   \bigg[2 \tr \bigg\{ \varphi\varphi\dagg 
       \bigg(FF\dagg + \1{4} \square  \varphi \varphi\dagg 
      + \1{4} \varphi \square  \varphi\dagg
    -\frac{1}{2} \partial_m \varphi \partial^m \varphi^\dagger 
      - {i \over 2} (\varphi \dellr_m \varphi\dagg)v^m \non
 && \hs{28}
  + \1{2}\varphi\varphi\dagg \big(D - \1{2} v^2\big) \bigg) \bigg\}
  - \frac{1}{2} \tr (i\varphi\dellr_m\varphi\dagg 
                     + \varphi\varphi\dagg v_m)^2
     \bigg] \, ,\non
&&    
 \tr (M_2)|_{\theta^2\thb^2} = \tr \bigg[
  -\1{2} \del_m Z\dagg \del^m Z 
   + \1{4} \square Z\dagg Z 
   + \1{4} Z\dagg \square  Z 
   - {i\over 2} (Z \dellr_m Z\dagg + Z\dagg \dellr_m Z) v^m 
  \non
&& \hs{20} 
   + \1{2} Z\dagg v_m Z v^m - \1{4} (Z\dagg Z + Z Z\dagg)v^2 
   - \1{2} (Z\dagg Z - Z Z\dagg)D + F_Z{}\dagg F_Z \bigg] \;.
 \label{comp.}
\eeq
with $A \dellr B \equiv A \del B - (\del A) B$.
Defining 
\beq
 X^a \equiv \tr (M_1{}^a) 
 \equiv C^a - \theta \sig^m \thb u^a_m 
 + \1{2} \theta^2\thb^2 E^a \;, 
\eeq
we can calculate an arbitrary function of them, 
to give~\cite{HN2}
\beq
 g(X) 
= 
 g(C) - \theta \sig^m \thb g,_a u^a_m 
  + \1{2} \theta^2 \thb^2 
   \left[ g,_a(C) E^a 
   - \1{2} g,_{ab} (C) u^a u^b \right]
 \label{g}
\eeq
Here $,_{a}$ denotes a differentiation with respect to 
$C^a$.
Substituting Eq.~(\ref{g}) with Eqs.~(\ref{comp.}) 
into the Lagrangian (\ref{deformed}), 
we obtain the component Lagrangian explicitly.

We have constructed the most general form of the 
effective Lagrangian for non-Abelian vortices 
which contains the deformation of the moduli metric 
compatible with SUSY and the global symmetry on the world-volume. 
We conjecture that the freedom of this deformation resolves 
the discrepancy between the Hanany-Tong metric and the 
Manton metric, although it remains to demonstrate 
this point explicitly.

\section{
Discussion}
\label{sc:discussion}

We have obtained a nonlinear sigma model on 
vortex world-volume 
formulated by auxiliary gauge fields. 
This fact raises a possibility that 
our model might provide an interesting possibility 
for dynamically induced gauge bosons~\cite{BKY} 
on the brane. 
Models of induced gauge bosons have been discussed 
in a number of different context previously
~\cite{Akama,DGS}. 
This type of composite gauge bosons can offer an 
alternative\cite{Ak}  
to the localization problem for the gauge bosons which 
has been notoriously difficult. 
In the case of a single vortex ($k=1$) we have 
the ${\bf C}P^N$ model formulated by an auxiliary 
$U(1)$ gauge field. 
It is well known that the large-$N$ analysis 
of the ${\bf C}P^N$ model displays 
a dynamically induced $U(1)$ gauge boson 
as a composite of Nambu-Goldstone bosons provided that 
the FI-term $c$ is proportional to $N$~\cite{DDL,BKY}. 
In our case, we have $c \sim 1/g^2 = N/ g^2 N$ 
with $g$ the gauge coupling of the original gauge theory. 
If one takes the limit of $N \to \infty$ with $g^2 N$ fixed, 
one has a dynamical $U(1)$ gauge boson on a single vortex.
So it is weak coupling limit in the original gauge theory.
In the case of multi-vortices we will have 
$U(k)$ gauge fields for coincident vortices 
and $U(1)^k$ gauge fields for separate vortices. 
It resembles to the situation of D-branes.
Thus we may have localized gauge fields on vortices induced by 
quantum effects. 
Let us note that this model gives 
the following interesting problem 
quantum mechanically. 
The ${\bf C}P^N$ model in four dimensions 
has the so-called sigma-model anomaly 
proportional to the first Chern class 
originated from the $U(1)$ gauge field~\cite{Moore:1984dc}. 
Presumably this is related with 
the gauge anomaly in six-dimensions 
proportional to\footnote{
We would like to thank David Tong for pointing this out.
}
 $N_{\rm F} - 2 N_{\rm C}$. 
We can add matters cancelling anomaly 
into the four-dimensional ${\bf C}P^N$ model 
by starting from the six-dimensional theory 
with $N_{\rm F} = 2 N_{\rm C}$.  
Investigating such a relation of quantum anomaly 
in four and six dimensions is 
a very interesting problem to be explored.

Another interesting possibility is the application to 
composite models of matter fields such as quarks. 
In four spacetime dimensions, 
SUSY nonlinear sigma models were 
proposed as models of composite quarks in which 
we identify superpartners of Nambu-Goldstone bosons as quarks~\cite{BLPY}. 
Phenomenological viability has been 
studied by working out various coset spaces. 
Previously not much attention have been payed as to 
the origin of the underlying SUSY linear theory which 
realizes the appropriate global symmetry that is broken 
spontaneously. 
Non-Abelian vortices may provide the origin of these models. 
We expect that such a problem can be considered realistically 
by introducing additional matter fields or 
taking different gauge group in the original gauge theory. 

Instead of the $U(N)$ gauge group which 
gives a coset space ${\bf C}P^{N-1}$ for a single vortex, 
we can consider other gauge groups. 
A single non-Abelian vortex in the theory with
$SO(N) \times U(1)$ gauge group and 
appropriate number of flavors 
is expected to give the effective theory with 
another coset space, 
the quadric surface 
$Q^{N-2} \equiv SO(N)/[SO(N-2)\times U(1)]$ 
which is also a Hermitian symmetric space. 
The sigma model Lagrangian on $Q^{N-2}$ was formulated 
in \cite{HN1,HKNT} 
as a $U(1)$ SUSY gauge theory 
by introducing a superpotential $W = \sigma \ph^a \ph^a$ 
$(a=1,\cdots,N)$ with
an auxiliary chiral superfield $\sigma$ with $U(1)$ 
charge $-2$. 
The equation of motion for $\sigma$ gives a constraint 
$\ph^a \ph^a = 0$ 
defining $Q^{N-2}$ embedded into ${\bf C}P^{N-1}$.
For $k$-vortices, the gauge group in the effective theory 
is extended to $U(k)$, and the matter chiral superfields 
$\varphi^{a}$ to a $k$ by $N$ matrix $\varphi^{Ia}$ ($I=1, \cdots, k$), 
similarly to sect.\ref{sc:sym-moduli-space}. 
The auxiliary chiral superfield $\sigma$ may then be 
extended to a $U(k)$ symmetric tensor $\sigma^{IJ}$, 
which appears in the superpotential\footnote{
We would like to thank David Tong for clarifying 
this point.
} 
as $W = \tr (\sig \ph \ph^T)$. 
Other Hermitian symmetric spaces, formulated 
in \cite{HN1} as SUSY gauge theories, may also be 
realized as the effective theory on a single vortex 
by taking other gauge groups with appropriate number of flavors 
in the original gauge theory.

In the case of $U(1)$ (Abelian) vortex, 
it was found that gauge field is localized 
by warp factors if we couple the theory 
with gravity~\cite{Gravity}. 
We expect that the same mechanism will work for 
non-Abelian vortices also.
Analysis for non-Abelian vortices in $D=6$ SUGRA remains 
as a future problem.

Both effective Lagrangians constructed 
in \cite{Hanany:2003hp} 
and by the Manton's method \cite{Ma} include 
up to two derivatives  
with respect to world-volume coordinates. 
They are valid provided zero-mode fluctuations 
vary only weakly on the world volume : 
$\del_m ({\rm fields}) << 1$. 
In the case of a single ANO vortex ($N=1, k=1$), 
the effective action 
including higher derivative corrections 
was constructed by using nonlinear realizations and 
the Green-Schwarz method~\cite{HLP}. 
They are supersymmetric generalization of the Nambu-Goto 
action. 
The action for non-Abelian vortices of the Nambu-Goto type 
is an interesting open problem.

\section*{Acknowledgements}
We would like to thank David Tong for a useful discussion. 
M.E. gratefully acknowledges 
support from the Iwanami Fujukai Foundation and 
from a 21st Century COE Program at 
Tokyo Tech ``Nanometer-Scale Quantum Physics" by the 
Ministry of Education, Culture, Sports, Science 
and Technology.
The work of M. N. is supported in part by JSPS. 
This work is supported in part by Grant-in-Aid for Scientific 
Research from the Ministry of Education, Culture, Sports, 
Science and Technology, Japan No.13640269 and 16028203 
for the priority area ``Origin of Mass'' (NS). 

\end{document}